\documentclass[apj,twocolumn,twocolappendix,numberedappendix]{openjournal}

\usepackage[dvipsnames]{xcolor}
\usepackage{hyperref}
\hypersetup{colorlinks=true,citecolor=blue,linkcolor=blue,urlcolor=blue,pdftitle={Estimating Cluster Galaxy Infall Time from Phase Space in TNG},pdfauthor={Cameron Lawlor-Forsyth},pdfsubject={}}
\usepackage{orcidlink}
\usepackage[hang]{footmisc}
\definecolor{doiPink}{RGB}{255, 0, 255}

\usepackage{enumitem}

\usepackage[largesc]{newtx}
\usepackage{couriers}
\DeclareSymbolFont{CMop}{OT1}{cmr}{m}{n}
\SetSymbolFont{CMop}{bold}{OT1}{cmr}{bx}{n}
\DeclareSymbolFont{CMlet}{OML}{cmm}{m}{it}
\SetSymbolFont{CMlet}{bold}{OML}{cmm}{b}{it}
\DeclareSymbolFont{CMSy}{OMS}{cmsy}{m}{n}
\SetSymbolFont{CMSy}{bold}{OMS}{cmsy}{b}{n}
\DeclareSymbolFont{AMSa}{U}{msa}{m}{n}
\SetSymbolFont{AMSa}{bold}{U}{msa}{m}{n}

\AtBeginDocument{%
    \DeclareMathSymbol{/}{\mathord}{CMop}{"2F}%
    \DeclareMathSymbol{+}{\mathbin}{CMop}{"2B}%
    \DeclareMathSymbol{-}{\mathbin}{CMSy}{"00}%
    \DeclareMathSymbol{=}{\mathrel}{CMop}{"3D}%
    \DeclareMathSymbol{<}{\mathrel}{CMlet}{"3C}%
    \DeclareMathSymbol{>}{\mathrel}{CMlet}{"3E}%
    \DeclareMathSymbol{\leqslant}{\mathrel}{AMSa}{"36}%
    \DeclareMathSymbol{\geqslant}{\mathrel}{AMSa}{"3E}%
    \DeclareMathSymbol{\lesssim}{\mathrel}{AMSa}{"2E}%
    \DeclareMathSymbol{\gtrsim}{\mathrel}{AMSa}{"26}%
    \DeclareMathSymbol{\sim}{\mathrel}{CMSy}{"18}%
    \DeclareMathSymbol{\pm}{\mathrel}{CMSy}{"06}%
    \DeclareMathSymbol{\approx}{\mathrel}{CMSy}{"19}%
    \DeclareMathSymbol{\times}{\mathbin}{CMSy}{"02}%
    \DeclareMathSymbol{\Delta}{\mathalpha}{CMop}{1}%
    \DeclareMathSymbol{\Omega}{\mathalpha}{CMop}{10}%
    \DeclareMathSymbol{\alpha}{\mathalpha}{CMlet}{11}%
    \DeclareMathSymbol{\beta}{\mathalpha}{CMlet}{12}%
    \DeclareMathSymbol{\delta}{\mathalpha}{CMlet}{14}%
    \DeclareMathSymbol{\mu}{\mathalpha}{CMlet}{22}%
    \DeclareMathSymbol{\sigma}{\mathalpha}{CMlet}{27}%
}

\makeatletter
    \patchcmd{\NAT@citex} 
        {\@citea\NAT@hyper@{%
            \NAT@nmfmt{\NAT@nm}%
            \hyper@natlinkbreak{\NAT@aysep\NAT@spacechar}{\@citeb\@extra@b@citeb}%
            \NAT@date}}
        {\@citea
        \NAT@nmfmt{\NAT@nm}%
        \NAT@aysep\NAT@spacechar
        \NAT@hyper@{\NAT@date}}%
        {}{}
    \patchcmd{\NAT@citex} 
        {\@citea\NAT@hyper@{%
            \NAT@nmfmt{\NAT@nm}%
            \hyper@natlinkbreak{\NAT@spacechar\NAT@@open\if*#1*\else#1\NAT@spacechar\fi}%
            {\@citeb\@extra@b@citeb}%
            \NAT@date}}
        {\@citea
        \NAT@nmfmt{\NAT@nm}%
        \NAT@spacechar\NAT@@open\if*#1*\else#1\NAT@spacechar\fi
        \NAT@hyper@{\NAT@date}}%
        {}{}
\makeatother

\newlength{\licenseiconwidth}
\newlength{\licenseicongap}
\newlength{\licenseindent}
\newcommand{\licensebox}{%
  \begin{figure}[!b]
    \footnotesize\linespread{1.2}\selectfont
    \setlength{\parindent}{0pt}
    \begingroup
      \parshape=3
        \licenseindent \dimexpr\columnwidth-\licenseindent\relax
        \licenseindent \dimexpr\columnwidth-\licenseindent\relax
        0pt \columnwidth
      \noindent
      \makebox[0pt][l]{%
        \hspace*{-\licenseindent}%
        \smash{%
          \raisebox{-1.1\baselineskip}[0pt][0pt]{%
            \includegraphics[width=\licenseiconwidth]{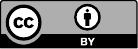}%
          }%
        }%
      }%
      Original content from this work may be used under the terms of the \href{https://creativecommons.org/licenses/by/4.0/}{Creative Commons Attribution 4.0 licence}. Any further distribution of this work must maintain attribution to the author(s) and the title of the work, journal citation and DOI.\par
    \endgroup
  \end{figure}
}

\journalinfo{The Open Journal of Astrophysics}
\historyline{Received 2026 July 22}
\shortauthors{Lawlor-Forsyth}

\begin{document}

\title{Estimating Cluster Galaxy Infall Time from Phase Space in TNG\vspace{-1.5cm}}

\author{Cameron~Lawlor-Forsyth$^{\text{\color{blue}1,2}}$\orcidlink{0000-0002-2958-0593}\vspace{0.1cm}}

\affiliation{$^{\text{1}}$Department of Physics and Astronomy, University of Waterloo, Waterloo, ON N2L 3G1, Canada; \href{mailto:clawlorforsyth@uwaterloo.ca}{clawlorforsyth@uwaterloo.ca}}
\affiliation{$^{\text{2}}$Waterloo Centre for Astrophysics, University of Waterloo, Waterloo, ON N2L 3G1, Canada}

\begin{abstract}
Using the TNG300 and TNG-Cluster simulations, phase space-based estimates for galaxy infall time onto clusters with $M_{\mathrm{h}} \geqslant 10^{14}~M_{\odot}$ are determined for all simulation snapshots with $z \leqslant 1$. These infall time estimates are consistent with previous work in the literature, and can be readily applied to observations of cluster galaxies, where no other estimate is feasible, based only on the galaxy's position in projected phase space. The analysis pipeline used to produce these infall time estimates is publicly released, as are the resulting data products, as well as a tutorial that outlines their use. It is anticipated that both the data products and the provided tutorial will facilitate future research within the community which require infall time estimates, and will provide an additional baseline for studies investigating the orbital histories of galaxies in dense environments like galaxy clusters.
\end{abstract}

\keywords{\href{http://astrothesaurus.org/uat/584}{Galaxy clusters (584)}; \href{http://astrothesaurus.org/uat/594}{Galaxy evolution (594)}; \href{http://astrothesaurus.org/uat/597}{Galaxy groups (597)}}

\maketitle

\section{Introduction}
\setcounter{footnote}{2}

\licensebox

The environment that a galaxy resides in directly influences its evolution. This has been well established over the past decades, and is seen most evidently via the morphology--density relation \citep[e.g.,][]{dressler1980}. Galaxies in dense environments are subject to additional physical effects that galaxies in the field do not experience, such as ram pressure stripping \citep[e.g.,][]{gunn1972}, strangulation \citep[e.g.,][]{larson1980}, and gravitational effects \citep[e.g.,][]{farouki1981}, all of which promote galaxy quenching.

Understanding the orbital history of a galaxy in dense environments such as galaxy clusters is therefore important to better understand galaxy evolution \citep[e.g.,][]{oman2013}. In particular, determining when a galaxy first fell into a galaxy cluster can clarify potential orbital histories, and thus potential evolutionary schemes \citep[e.g.,][]{oman2016}. Unfortunately, determining cluster infall time is not feasible observationally. However, cosmological galaxy simulations provide an attractive testbed for determining cluster infall times for simulated galaxies \citep[e.g.,][]{dou2025,masson2026}, and their results can be used to infer infall times for samples of observed galaxies.

Over the past decade, much work has been done to estimate galaxy infall time based on position in projected phase space \citep[e.g.,][]{oman2013,oman2016,rhee2017,pasquali2019}. Recently, \citet{dou2025} completed an analysis using the IllustrisTNG simulations \citep{nelson2018,pillepich2018b,springel2018} to determine galaxy infall times onto clusters for galaxies at $z = 0$. Here, their method is expanded upon to all snapshots with $z \leqslant 1$. Importantly, the code is also made available to reproduce this analysis, the end data products, and a tutorial that uses the data to infer infall times for samples of observed galaxies, based on position in projected phase space.
 
This paper is structured as follows: in Section~\ref{sec:data}, the IllustrisTNG simulations are described, and in particular the TNG-Cluster follow-up simulations which have recently become available. In Section~\ref{sec:method}, the methodology to determine galaxy infall times from phase space is described, highlighting the extension to higher redshifts. Finally, in Section~\ref{sec:results}, the results and concluding remarks are presented. The data products presented here, alongside the accompanying tutorial, are intended to serve as a resource for the community, and to facilitate analyses relying on infall time estimates.

Throughout this paper a flat $\Lambda$CDM cosmology that is consistent with the TNG simulation \citep{weinberger2017,pillepich2018a} is adopted, based on the Planck intermediate results \citep{planck2016}: $H_{0} = 67.74~\text{km s}^{-1}~\text{Mpc}^{-1}$, $\Omega_{\text{m}} = 0.3089$, $\Omega_{\Lambda} = 0.6911$, and $\Omega_{\text{b}} = 0.0486$.

\section{IllustrisTNG}\label{sec:data}

The IllustrisTNG simulations\footnote{\href{https://www.tng-project.org}{https:$//$www.tng-project.org}} \citep[e.g.,][]{nelson2018,pillepich2018b,springel2018} are the successor to the original Illustris simulation \citep{genel2014,vogelsberger2014a,vogelsberger2014b}, and are cosmological gravo-magneto-hydrodynamical simulations with a revised galaxy formation model \citep{weinberger2017,pillepich2018a} that builds on that used in Illustris \citep{vogelsberger2013,torrey2014}. Using a cosmology consistent with the \citet{planck2016} results, IllustrisTNG employs the moving-mesh code \textsc{arepo} \citep[e.g.,]{springel2010}, and draws stellar populations from the \citet{chabrier2003} initial mass function \citep{pillepich2018a}. As detailed in \citet{rodriguez2019}, the IllustrisTNG model aims to reproduce several observable phenomena at $z = 0$, including the galaxy mass function and the stellar mass--halo mass relation. All simulation data from IllustrisTNG are publicly available \citep{nelson2019a}.

As stated above, the TNG300-1 \citep[hereafter TNG300;][]{nelson2019a} and the follow-up TNG-Cluster \citep{nelson2024} are used in the analysis. TNG300 contains a cubic volume of $205 h^{-1} \approx 302.6~\text{Mpc}$ on a side, and follows the evolution of $2 \times 2500^{3}$ resolution elements (dark matter and baryons), with an average baryon particle mass of $1.1 \times 10^{7}~M_{\odot}$ \citep[][]{nelson2018,pillepich2018b,springel2018}. The critical quantity is the number of galaxy clusters that the TNG300 simulation contains, which numbers 239 clusters with $10^{14.0} \leqslant M_{\text{halo}}/M_{\odot} \leqslant 10^{14.5}$ at $z = 0$ \citep{pillepich2018b,nelson2024}, but that shows a steep drop off in the halo mass function beyond $M_{\text{halo}} \sim 10^{14.6}~M_{\odot}$ \citep{pillepich2018b,nelson2024}. Indeed, in TNG300, there exist only 41 clusters with $M_{\text{halo}} \geqslant 10^{14.5}~M_{\odot}$, and of these, only three have $M_{\text{halo}} \geqslant 10^{15.0}~M_{\odot}$ \citep{pillepich2018b,nelson2024}. In total, there are 280 clusters with $M_{\text{halo}} \geqslant 10^{14.0}~M_{\odot}$ in TNG300 at $z = 0$ \citep{pillepich2018b,nelson2024}.

To better sample the high mass end of the halo mass function, the recently available follow-up TNG-Cluster\footnote{\href{https://www.tng-project.org/cluster}{https:$//$www.tng-project.org$/$cluster}} simulation \citep{nelson2024} is additionally used. The objective of the TNG-Cluster simulation is to substantially increase the statistical sampling of massive galaxy clusters with $M_{\text{halo}} \gtrsim 10^{15.0}~M_{\odot}$ \citep{nelson2024}. It does this by encompassing a cubic volume of $680 h^{-1} \approx 1003.8~\text{Mpc}$ on a side, following the evolution of $2 \times 8192^{3}$ resolution elements\footnote{The quoted number of resolution elements represents the effective full-volume equivalent \citep{nelson2024}.} (dark matter and baryons), with an average baryon particle mass of $1.2 \times 10^{7}~M_{\odot}$ \citep[thereby matching the resolution of TNG300;][]{nelson2024}, while maintaining the IllustrisTNG galaxy model \citep{weinberger2017,pillepich2018a} and cosmology \citep{nelson2024}. Critically, TNG-Cluster contains 356 clusters with $M_{\text{halo}} \geqslant 10^{14.0}~M_{\odot}$ at $z = 0$, where 207 have $10^{14.5} \leqslant M_{\text{halo}}/M_{\odot} \leqslant 10^{15.0}$, and 92 are more massive than $M_{\text{halo}} = 10^{15.0}~M_{\odot}$ \citep{nelson2024}. Though TNG-Cluster technically re-simulates 352 clusters drawn from a much larger $1~\text{Gpc}^{3}$ volume \citep{nelson2024}, in practice, the results from the TNG-Cluster simulation can be analyzed analogously to those from TNG300. By combining these two large volume simulations, TNG300 and TNG-Cluster, the halo mass distribution becomes flattened from $M_{\text{halo}} = 10^{14.0}~M_{\odot}$ to $M_{\text{halo}} = 10^{15.1}~M_{\odot}$, showing a steep drop off after $M_{\text{halo}} \sim 10^{15.2}~M_{\odot}$ \citep{nelson2024}. In total, there exist 1281 simulated clusters at $z = 0$ with $M_{\text{halo}} \geqslant 10^{14.0}~M_{\odot}$. All snapshots with $z \leqslant 1$ are additionally considered, which similarly contain hundreds of massive clusters. In particular, with the combination of TNG300 and TNG-Cluster, there exist 514 clusters at $z = 1$ with $M_{\text{halo}} \geqslant 10^{14.0}~M_{\odot}$.

By using all snapshots with $z \leqslant 1$ in the simulation, the orbital histories of thousands of simulated galaxies are tracked, thereby robustly sampling the available phase space. Though \citet{dou2025} and other authors refer to ``phase space'' when considering the full three-dimensional components of their simulated galaxies, and use ``projected phase space'' for their two-dimensional counterparts, here ``phase space'' is used to discuss the projected two-dimensional version for simplicity.

\section{Methodology}\label{sec:method}

The method of \citet{dou2025} is built upon. \citet{dou2025} determined phase space-based infall time estimates for galaxies in TNG300, but exclusively at $z = 0$. Here, their method is extended to all snapshots with $z \leqslant 1$, using both TNG300 and TNG-Cluster. The following describes the methodology for a single snapshot.

Using the $z = 0$ snapshot as an example, simulated clusters with $M_{\text{halo}} \geqslant 10^{14}~M_{\odot}$ are selected. A requirement that all cluster galaxies must reside within $3 R_{200}$ for a cluster to be considered valid is then implemented, in order to reduce the impact of cluster mergers on the orbital histories of the galaxies \citep{dou2025}. The central galaxies (i.e. brightest cluster galaxies) embedded within the halo are then used as proxies to track the positions and motions of the clusters, where these central galaxies almost always remain at the center of a given cluster. Any clusters where the central galaxy is misclassified in more than three consecutive snapshots, as can sometimes happen at high redshift, are further excluded. These misclassifications may be related to the ``subhalo switching problem'' \citep{rodriguez2015}. For all remaining clusters, satellite galaxies associated with each cluster are then selected, removing central galaxies, as well as those that passed through a cluster and subsequently exited (i.e. temporary infall galaxies). In order to match the analysis of \citet{dou2025}, galaxies with $M_{*} \geqslant 10^{9}~M_{\odot}$ are used, to ensure that they are well resolved, consistent with \citet{pillepich2018b}.

The phase space diagram is then constructed by aligning the line-of-sight direction with each of the simulation coordinate axes, thereby tripling the sample size. The line-of-sight velocity is simply the velocity component along the line of sight, while the projected radius, $R_{\text{2D}}$, is the Euclidean distance from a given galaxy to the central galaxy on the projected plane \citep{dou2025}. Individual clusters are stacked together after normalizing the projected radii by $R_{200}$, and the line-of-sight velocities by the line-of-sight velocity dispersions, $\sigma_{\text{LoS}}$. A Hubble flow correction is additionally applied along the line of sight using the components of the cluster-centric velocity and radial vectors. The merger trees for each galaxy are then leveraged, tracking back along the main branch of the progenitor tree \citep[e.g.,][]{genel2018,nelson2019b} to determine when a galaxy first fell into a cluster. Matching \citet{dou2025}, the infall time, $t_{\text{infall}}$, is defined as the lookback time when a galaxy first crosses $3 R_{200}$ of the cluster. After stacking the clusters, median infall times based on location in phase space are then determined. As noted by \citet{masson2026}, the resulting infall time distributions are complex, but median values nonetheless provide a baseline that is otherwise inaccessible with observations.

\begin{figure*}[t]
    \centering
    \includegraphics[width=\textwidth]{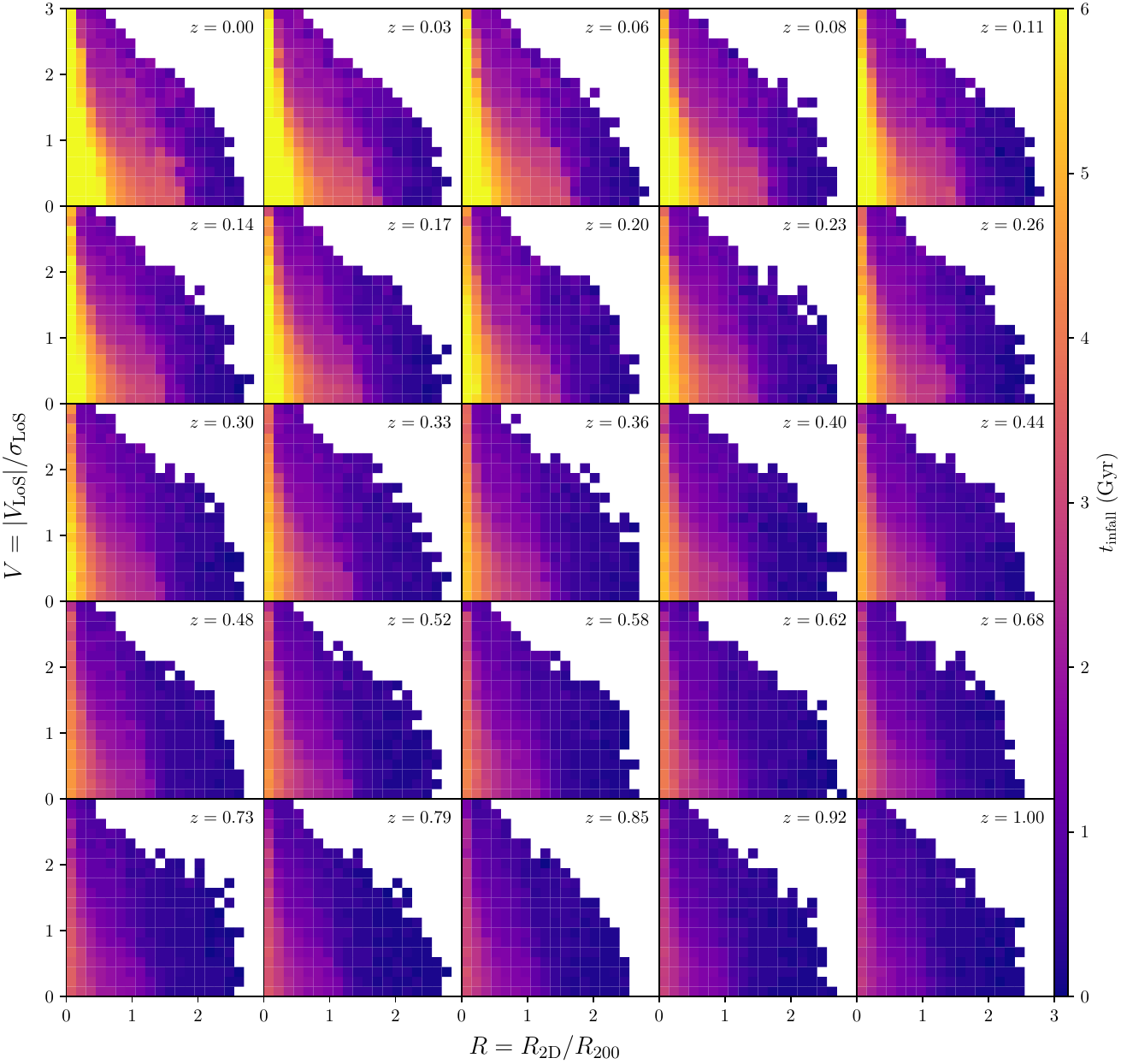}
    \caption{Phase space diagrams showing the median infall time (in Gyr) per bin for redshift $z = 0$ (snapshot 99) to $z = 1$ (snapshot 50). In each panel, the corresponding redshift is shown in the upper right corner, and the phase space bins are color-coded according to median infall time.}
    \label{fig:tinfall}
\end{figure*}

This procedure for all snapshots with $z \leqslant 1$ is then repeated, building up a stack of phase space maps. By considering the full stack of phase space maps, infall time estimates can be estimated through interpolation. In practice, bilinear interpolation is used when considering the phase space at a single snapshot. However, when considering the full stack of phase space maps, the median infall time of a galaxy can be determined using
\begin{equation}
    t_{\text{infall}} = t_{\text{median}} (R, V, t_{\text{lb}}),
\end{equation}
where $t_{\text{median}}$ is the median infall time, and is a function of (normalized) position in phase space, $R = R_{\text{2D}}/R_{200}$ and $V = |V_{\text{LoS}}|/\sigma_{\text{LoS}}$, as well as the lookback time, $t_{\text{lb}}$, of the redshift (i.e. snapshot) under consideration. Correspondingly, trilinear interpolation is used to determine $t_{\text{infall}}$ after accounting for lookback time. Through this process, from $z = 1$ to $z = 0$, more than 350 simulated clusters are used per snapshot, with ${>} 20,000$ unique simulated galaxies per snapshot, in order to determine median infall times.

Beyond estimating the median infall time, the method of \citet{dou2025} can similarly be followed to determine the associated root mean square error (RMSE) for the infall times. The extensions to higher redshifts are likewise adopted as described.

\begin{figure*}[t]
    \centering
    \includegraphics[width=\textwidth]{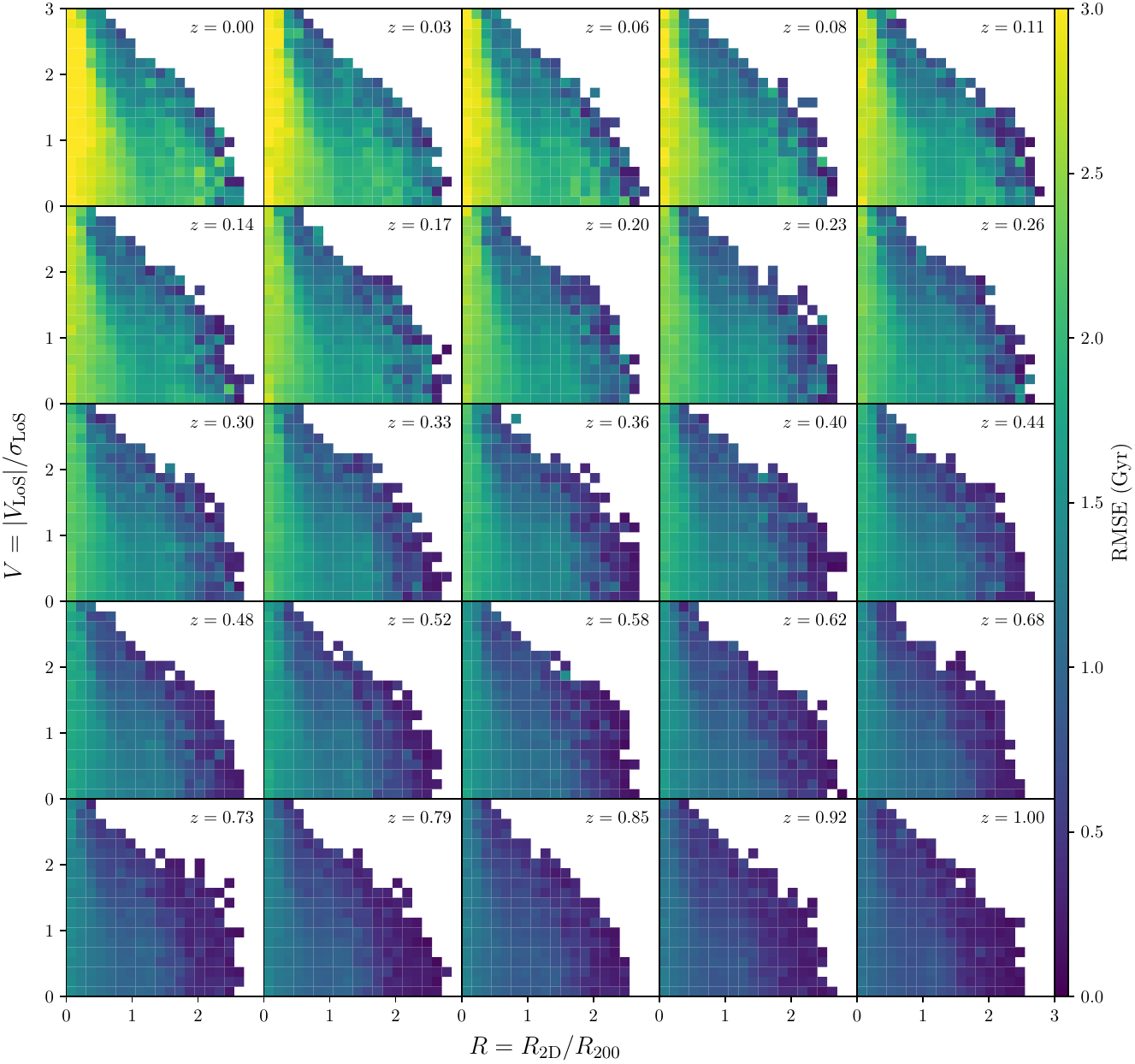}
    \caption{Phase space diagrams showing the root mean square error (RMSE, in Gyr) per bin for redshift $z = 0$ (snapshot 99) to $z = 1$ (snapshot 50). In each panel, the corresponding redshift is shown in the upper right corner, and the phase space bins are color-coded according to root mean square error.}
    \label{fig:RMSE}
\end{figure*}

\section{Results and Conclusion}\label{sec:results}

In Figure~\ref{fig:tinfall}, the resulting phase space diagrams for $z 
\leqslant 1$ (snapshots 50--99) are shown, where the median infall time is given in Gyr, and the phase space bins are color-coded according to median infall time. Though only every second snapshot is shown for conciseness, all snapshots are available in the available data products.

At higher redshifts, $z \sim 1$, infall times are modest, commonly taking values ${<} 2~\text{Gyr}$, with a gradient toward the inner regions of the cluster, where infall times are largest. Moving to lower redshift, this situation evolves, such that by $z = 0$, there exist a population of galaxies with ancient infall times ($t_{\mathrm{infall}} > 6~\text{Gyr}$) near the cluster core. Median infall times likewise show a gradient to lower values when extending out past the cluster core at low redshift. This situation is consistent with \citet{dou2025} when considering $z = 0$ alone (e.g., see their Figure~2), and also with \citet{masson2026} when considering specific redshifts (e.g., see their Figure~2).

In Figure~\ref{fig:RMSE}, the resulting phase space diagrams for RMSE are shown, and are arranged as in Figure~\ref{fig:tinfall}. The phase space bins are color-coded according to RMSE, which is given in Gyr. Similar to the situation for median infall time, the RMSE maps evolve from high redshift to low redshift, where at $z = 0$, RMSE values can reach ${\sim} 3~\text{Gyr}$ near the cluster core, indicating that there is intrinsic scatter about the median infall time for these galaxies. These results are likewise consistent with \citet{dou2025} at $z = 0$ (e.g., see their Figure~2), as well as \citet{masson2026} for their specified redshifts (e.g., see their Figure~11).

From these figures, it is clear that the median infall time and root mean square error evolve from $z = 1$ to $z = 0$. Maps such as those presented here can therefore be useful for estimating infall times for observational samples of galaxies, where no other determination is feasible, unlike in simulations. The maps presented here are consistent with other recent works in the literature \citep[e.g.,][]{dou2025,masson2026}, and can be directly applied to observational samples.

To facilitate this, the analysis pipeline, the end data products, and a tutorial that uses the data to infer infall times is publicly released. The full pipeline, including the data, analysis methods, and tutorial, is available at \href{https://doi.org/10.5281/zenodo.21496572}{https:$//$doi.org$/$10.5281$/$zenodo.21496572}, while the analysis pipeline and tutorial are additionally available at \href{https://github.com/camlawlorforsyth/tng-galaxy-infall-times}{https:$//$github.com$/$camlawlorforsyth$/$tng-galaxy-infall-times}. This suite of data products and the accompanying tutorial aim to lower the barrier to entry for this type of analysis, and to support ongoing community initiatives. Our analysis, as well as our data products, should similarly be complementary to other recent work in the literature, including \citet{masson2026}. It is hoped that the community will benefit from the data products and the tutorial introduced in this work.

\section*{Acknowledgments}

The IllustrisTNG simulations were undertaken with compute time awarded by the Gauss Centre for Supercomputing (GCS) under GCS Large-Scale Projects GCS-ILLU and GCS-DWAR on the GCS share of the supercomputer Hazel Hen at the High Performance Computing Center Stuttgart (HLRS), as well as on the machines of the Max Planck Computing and Data Facility (MPCDF) in Garching, Germany. The TNG-Cluster simulation has been executed on several computer clusters: with compute time awarded under the TNG-Cluster project on the HoreKa supercomputer, funded by the Ministry of Science, Research and the Arts Baden-Württemberg and by the Federal Ministry of Education and Research; the bwForCluster Helix supercomputer, supported by the state of Baden-Württemberg through bwHPC and the German Research Foundation (DFG) through grant INST 35$/$1597-1~FUGG; the Vera cluster of the Max Planck Institute for Astronomy (MPIA), as well as the Cobra and Raven clusters, all three operated by the MPCDF; and the BinAC cluster, supported by the High Performance and Cloud Computing Group at the Zentrum für Datenverarbeitung of the University of Tübingen, the state of Baden-Württemberg through bwHPC and the German Research Foundation (DFG) through grant no INST 37$/$935-1~FUGG.

This research has made use of the Astrophysics Data System, funded by NASA under Cooperative Agreement 80NSSC25M7105, as well as \texttt{Astropy},\footnote{\href{https://www.astropy.org}{https:$//$www.astropy.org}} a community-developed core Python package and an ecosystem of tools and resources for astronomy \citep{astropy2013,astropy2018,astropy2022}.

\vspace{0.3cm}
\textit{Software:} \texttt{Astropy} \citep{astropy2013,astropy2018,astropy2022}, \texttt{h5py} \citep{collette2023}, \texttt{Matplotlib} \citep{hunter2007}, \texttt{NumPy} \citep{harris2020}, \texttt{SciPy} \citep{virtanen2020}

\section*{ORCID iDs}
\begingroup
\raggedright
Cameron~Lawlor-Forsyth \orcidlink{0000-0002-2958-0593} \\\href{https://orcid.org/0000-0002-2958-0593}{https:$//$orcid.org$/$0000-0002-2958-0593}\par
\endgroup

\bibliography{references}{}
\bibliographystyle{aasjournal}

\end{document}